\documentclass[11pt]{article}
\newcommand{\pdr}{\partial}
\newcommand{\beq}{\begin{equation}}
\newcommand{\eeq}{\end{equation}}
\newcommand{\beqs}{\begin{eqnarray}}
\newcommand{\eeqs}{\end{eqnarray}}
\newcommand{\half}{\frac{1}{2}}
\newcommand{\ov}[1]{\frac{1}{#1}}
\def\del{\delta}
\def\eps{\epsilon}
\newcommand{\sech}{\mathrm{sech} \,}

\usepackage{appendix}
\usepackage{amsmath,amssymb}
\usepackage{graphicx}
\usepackage{color}
\usepackage[pdftex]{hyperref}
\usepackage{txfonts}
\usepackage{subfigure} 

\begin{document}

\title{\normalsize 
\hfill {\tt arXiv:1109.3745 [nlin.PS]} \\
\vskip 0.1mm 
\large \bf
A KdV-like advection-dispersion equation with some remarkable properties }

\author{\sc Abhijit Sen$^{1}$, Dilip P. Ahalpara$^{1}$ , Anantanarayanan Thyagaraja$^{2}$ and \\ \sc Govind S. Krishnaswami$^{3}$
\\ \\
$^{1}$Institute for Plasma Research, Bhat, Gandhinagar 382 428, India, \\ $^{2}$ H.H. Wills Physics Laboratory, University of Bristol, Tyndall Avenue, BS8 1TL, UK, \\ $^{3}$Chennai Mathematical Institute, SIPCOT IT Park, Siruseri 603103, India.
\\ \\
Email: {\tt senabhijit@gmail.com, dpa1951@gmail.com}, \\ {\tt a.thyagaraja@bristol.ac.uk \& govind@cmi.ac.in}}

\date{February 25, 2012}

\maketitle

\begin{abstract}
We discuss a new non-linear PDE,  $u_t + (2 u_{xx}/u) u_x = \epsilon u_{xxx}$, invariant under scaling of dependent variable and referred to here as SIdV. It is one of the simplest such translation and space-time reflection-symmetric first order advection-dispersion equations. This PDE (with dispersion coefficient unity) was discovered in a genetic programming search for equations sharing the KdV solitary wave solution. It provides a bridge between non-linear advection, diffusion and dispersion. Special cases include the mKdV and linear dispersive equations. We identify two conservation laws, though initial investigations indicate that SIdV does not follow from a polynomial Lagrangian of the KdV sort. Nevertheless, it possesses solitary and periodic travelling waves. Moreover, numerical simulations reveal recurrence properties usually associated with integrable systems. KdV and SIdV are the simplest in an infinite dimensional family of equations sharing the KdV solitary wave. SIdV and its generalizations may serve as a testing ground for numerical and analytical techniques and be a rich source for further explorations.
\end{abstract}


\vspace{1cm}

\centerline{\em Communications in Nonlinear Science and Numerical Simulation 17 (2012), pp. 4115-4124}

\thispagestyle{empty}

\clearpage


\footnotesize

\tableofcontents

\normalsize

\section{Introduction}

Several equations of physics have been discovered by searching for one that admits a particular type of solution that was known to exist on physical grounds. Perhaps the most celebrated example is the discovery of Schr\"{o}dinger's equation of quantum mechanics via de Broglie's hypothesis that free particles are described by plane matter waves. The KdV equation of fluid flow in a canal \cite{KdV},
	\beq
	u_t + 6 u u_x + u_{xxx} = 0
	\label{e:KdV-eqn}
	\eeq
was motivated (and in part discovered) by the search for an equation possessing Russell's \cite{scott-russell} `wave of translation' as a solution. This is the solitary wave
	\beq
	u(x,t) = \frac{c}{2} \sech^2\left(\frac{\sqrt{c}}{2} (x-ct - x_0) \right) \quad {\rm for} \quad c>0, \; x_0 \in \mathbb{R}.
	\label{e:KdV-solitary-wave-rtmov-elevation}
	\eeq
The KdV equation is the simplest conservative $1$-dimensional wave equation with weak advective non-linearity and dispersion. Thus it is widely applicable and has been used to model acoustic solitons in plasmas \cite{zabusky-kruskal,washimi-taniuti}, internal gravity waves \cite{benney} in the oceans and even blood pressure pulses \cite{yomosa}. Its significance was greatly amplified by Zabusky \& Kruskal's discovery \cite{zabusky-kruskal} that KdV displays the Fermi-Pasta-Ulam recurrent behavior and lack of thermalization \cite{fpu} in a spatially periodic domain (see also \cite{thyagaraja1979,thyagaraja1983}). This was subsequently attributed to the even more remarkable asymptotic superposition principle for scattering of KdV solitary waves and existence of infinitely many local conserved quantities in involution \cite{miura-gardner-kruskal}. Moreover, KdV is the prototype for an integrable non-linear PDE, its initial value problem can be solved by the inverse scattering transform \cite{miura-survey-IST}.

We were therefore excited to find that the KdV equation is not the only one with the $\sech^2$ solitary wave solution. This serendipitous discovery happened in the course of an investigation undertaken by two of the present authors \cite{abhijit-dilip} to improve the efficiency and accuracy of a genetic programming (GP) based  method (see \ref{a:GP-sniffer}) to deduce model equations from a known analytic solution \cite{kudryasov1}.  As a benchmark exercise to test out the method for application to nonlinear PDEs, the travelling wave (\ref{e:KdV-solitary-wave-rtmov-elevation}) was given to the program, expecting it to find the KdV equation. But, surprisingly, before finding the KdV equation, it found a {\em different} equation
	\beq
	u_t + \left( \frac{2 u_{xx}}{u} \right) u_x = u_{xxx}
	\label{e:AD-eqn}
	\eeq
which had the {\em same} solitary wave solution. Subsequently, we found that (\ref{e:AD-eqn}) is the simplest in a vast family of equations sharing the KdV solitary wave. We think of (\ref{e:AD-eqn}) as a non-linear wave equation for the dispersive advection of the real wave amplitude $u$. Unlike the KdV equation, where the advecting velocity $V = 6u$ is linear, here it is a quotient $V = (2 u_{xx} / u)$. As a consequence (\ref{e:AD-eqn}) is `scale-invariant' under dilation $u \to \lambda u$ of {\em dependent} variable. Scale-invariant advective velocities have appeared before e.g., the $\frac{{\bf E} \times {\bf B}}{{\bf B}^2}$ velocity that is invariant under a rescaling of fields, charges and currents in a plasma. We refer to (\ref{e:AD-eqn}) as the ($\eps =1$ case of the) SIdV equation\footnote{The acronym SIdV highlighting scale-invariance is related to KdV as Sine-Gordon is related to Klein-Gordon.}. Like KdV, faster SIdV solitary waves are narrower, but due to scale invariance, height and speed are generally unrelated, as in classical linear wave equations, but unlike KdV and the non-linear Schr\"odinger equation (NLSE).

In (\ref{e:AD-eqn}) the dimension-$L^3/T$ coefficients of the dispersive and advective terms are {\em both} equal to one. This is of course very special. More generally we consider the SIdV equation
	\beq
	u_t + \left( \frac{2 a u_{xx}}{u} \right) u_x = \eps \, a \: u_{xxx}.
	\label{e:sidv-eps}
	\eeq
$x \to \sqrt[3]{a} x$ eliminates the $L^3/T$-dimensional constant $a$, leaving one dimensionless parameter $\eps$ measuring the strength of dispersion relative to advection. Unlike KdV, where they scale differently in $x$, here both scale as $L^{-3}$. Though we have not found an experimental system modelled by (\ref{e:sidv-eps}), it is just as simple and universal among scale-invariant advection-dispersion equations as KdV is among all such equations. For generic $\eps$, we have identified $2$ conserved densities. This is similar to inviscid Eulerian hydrodynamics, but unlike KdV and NLSE which possess an infinite number. While SIdV shares solitary waves with KdV at $\eps = 1$, non-linear diffusion and solvability emerge elsewhere. Despite being non-linear, scale-invariance ensures SIdV has exact plane wave solutions. Furthermore, it possesses bounded spatially periodic travelling waves and similarity solutions. However, SIdV cannot arise from a polynomial Lagrangian in the sort of variable that works for KdV. We evade this obstruction at some special values of $\eps$. Remarkably, numerical evolution of (\ref{e:AD-eqn}) shows Fermi-Pasta-Ulam-like Birkhoff recurrence \cite{thyagaraja1983} despite no sign of soliton scattering!

\section{General properties of the SIdV equation}
\label{s:general-properties}

\subsection{Symmetries and Conservation laws}

SIdV (\ref{e:sidv-eps}) is a non-linear advection-dispersion equation for the real wave amplitude $u(x,t)$ that is being advected by the flow $V = 2u_{xx}/u$. It is scale-invariant\footnote{SIdV (\ref{e:sidv-eps}) is {\em not} invariant under dilations of $x$ and has a {\em dimensional} scale $a$ which we set to $1$.} under $u \to \lambda u$. In fact, (\ref{e:sidv-eps}) is the simplest {\em non-linear} translation and scale-invariant advection-dispersion equation that is first order in time. The lowest order dispersive term is $u_{xxx}$, so any such equation can be written as $u_t + V u_x = a \eps u_{xxx}$ for some scale-invariant advective velocity $V$. Without requiring scale-invariance, the simplest choice $V \propto u$ leads to KdV. KdV is symmetric under space-time (PT) reflection $(x,t) \to (-x,-t)$. Now if we also require scale-invariance, the simplest advecting velocities that preserve PT symmetry are $V \propto \frac{u_{xx}}{u}$ and $\frac{u}{u_{xx}}$. The former leads to (\ref{e:sidv-eps}) in units where $a=1$. The latter choice too has some notable properties (see \S\ref{s:discussion}).

Non-zero constants are the simplest solutions of (\ref{e:sidv-eps}). Linearization about a constant yields a plane wave with cubic dispersion $\omega = \eps k^3$ and the characteristic property that the ratio of phase to group velocity is $1/3$, as in the linear KdV equation. Remarkably, despite being non-linear, (\ref{e:sidv-eps}) also admits {\em exact} plane wave solutions $u = A \sin\left(kx - (\eps -2)k^3 t + \varphi \right)$.

KdV is invariant under Galilean boosts $x \to x + ct$ if $u$ transforms as $u \to u(x+ct,t) - (c/6)$. However, SIdV is not invariant under $\del u = b + c t u_x$ for any constants $b$ and $c \ne 0$. However, Galilean boosts could be implemented in a more intricate manner that we have not identified.

Multiplying by $u$, and using $uu_{3x} = \left(uu_{xx} - \half u_x^2 \right)_x$, (\ref{e:sidv-eps}) is written in conservation form
	\beq
	\half \left[ u^2 \right]_t + \left[ \left(1 + \frac{\eps}{2}\right) u_x^2 - \eps u u_{xx} \right]_x =0.
	\label{e:cons-law-form-sidv}
	\eeq
It follows that $I = \int u^2 \; \text{d}x$ is conserved. This suggests $u^2$ is the concentration of some substance whose total amount is conserved. Similarly, multiplying (\ref{e:sidv-eps}) by $u^{-2/\eps}$ we get
	\beq
	\left( u^{1-2/\eps} \right)_t + (2 - \eps) \left(u^{-2/\eps} u_{xx} \right)_x = 0.
	\label{e:cons-law-form-sidv-second}
	\eeq
So $J = \int u^{1-2/\eps} \: \text{d}x$ is also conserved. For e.g., when $\eps = 1$, $J = \int \ov{u} \: \text{d}x$, so on a bounded domain, $J$ is finite for any strictly positive/negative initial condition. These integrals of motion, travelling waves (\S \ref{s:trav-waves}) and numerical evolution (\S \ref{s:numerical-evolution-recurrence}) indicate that SIdV is generically non-dissipative.

\subsection{Preservation of positivity of $u(x)$}

At first sight, it appears that $u=0$ is a singular point of (\ref{e:sidv-eps}). But $u$ can vanish at points where $u_x$ or $u_{xx}$ also vanish, provided $u_{xx} u_x/u$ is finite, e.g. the above plane wave vanishes at isolated points. Among travelling waves, this is generic, near one where $u$ and $u_{xx}$ have common zeros, there is another solution with the same property. However, in the numerical and analytical examples studied, if $u(x,0) > 0$, it remains positive for $t > 0$. Let us use the second conserved quantity $J = \int \ov{u} \: \text{d}x$ to sketch why this is the case for $\eps=1$ on a bounded domain\footnote{A similar argument can be given for any $\eps$ for which a conserved density diverges at $u = 0$ sufficiently fast.}. Suppose $u(x,0)>0$ and $J$ is finite at $t=0$. At $t_1 > 0$, let $u(x)$ develop its first zero, this cannot be a first-order zero as $u$ was strictly positive. Then $J(t_1) = \infty$, contradicting the constancy of $J$! So strictly positive initial data cannot develop a zero and therefore must remain positive\footnote{We assume that $u(x)$ cannot develop an `integrable' zero for which $J$ is finite. An `integrable' zero where $u \sim |x|^a$ for $a < 1$ would mean $u$ forms a cusp and ceases to be thrice continuously differentiable. Assuming the solution remains sufficiently smooth, such possibilities are eliminated. Our numerical simulations did not indicate cusp formation, though it is an open question whether SIdV preserves regularity of initial data.}.

\subsection{Behaviour of SIdV at some special values of dispersion coefficient}
\label{s:behaviour-at-special-eps}

\subsubsection{Reduction to a linear dispersive wave equation when $\eps = -2/3$}
\label{s:eps-minus-2by3}

For $\eps = -2/3$, SIdV may be reduced to a {\em linear} dispersive equation for $u^2$. If we write $uu_{xx} = (uu_x)_x - u_x^2 = \half (u^2)_x - u_x^2$ in the conservation law form (\ref{e:cons-law-form-sidv}) of SIdV, we get
	\beq
	\half \pdr_t u^2 + \pdr_x \left[ \left(1 + \frac{3\eps}{2} \right) u_x^2 - \frac{\eps}{2} \left(u^2 \right)_{xx} \right] = 0.
	\eeq
If $\eps = -2/3$, we see that $\rho = u^2$ satisfies a linear KdV equation
	\beq
	\rho_t + \frac{2}{3} \rho_{xxx} = 0.
	\eeq
So for $\eps = -\frac{2}{3}$, $\rho = u^2$ is a sort of Cole-Hopf transformation that linearizes the equation. The general solution is a linear combination of plane waves
	\beq
	\rho(x,t) = u^2(x,t) = \int_{\mathbb{R}} \tilde \rho(k) \; e^{i\left(kx + \frac{2}{3} k^3 t \right)} \; \frac{\text{d}k}{2\pi} \quad \text{where} \quad
	\tilde \rho(k) = \int_{\mathbb{R}} u^2(x,0) \: e^{-ikx} \; \text{d}x.
	\eeq
This solution could serve as the $0^{\rm th}$ order of a perturbative solution of SIdV for nearby $\eps$.

\subsubsection{The dispersionless limit $\eps = 0$}
\label{s:dispersionless-case}

The special case $\eps =0$ of (\ref{e:sidv-eps}) gives a non-dispersive non-linear advection equation
	\beq
	u_t + \frac{2u_{xx}}{u} u_x = 0.
	\label{e:dispersionless-sidv}
	\eeq
It may be written as a conservation law $(u^2)_t + \left(2 u_x^2 \right)_x =0$ for the `charge' density $u^2$ with flux $u_x^2$. Being a second order parabolic PDE, (\ref{e:dispersionless-sidv}) may also be viewed as an unusual non-linear diffusion equation $u_t = \alpha u_{xx}$ for the `temperature' $u$. The effective thermal diffusivity $\alpha = -2u_x/u$ could be of either sign. Thus SIdV is a remarkable bridge connecting dispersion, non-linear advection and non-linear diffusion. We expect (\ref{e:dispersionless-sidv}) to have instabilities if $\alpha$ becomes negative since time-reversed heat equations are ill-posed. So we may think of the dispersive term in SIdV as a regularization of (\ref{e:dispersionless-sidv}), just as KdV is a regularization of the kinematic wave equation (KWE) $u_t + 6 u u_x=0$. Remarkably, even without a dispersive regularization, (\ref{e:dispersionless-sidv}) has smooth solutions. Indeed, unlike the KWE, (\ref{e:dispersionless-sidv}) admits waves that preserve their shape. The general travelling wave of (\ref{e:dispersionless-sidv}) is $u = A \cos (kx + 2 k^3 t + \phi)$. These plane waves may however be unstable, as the diffusivity oscillates in sign!

\subsubsection{SIdV to mKdV when $\eps = 2/3$}
\label{s:SIdV-to-mKdV}

We did not find any value of $\eps$ at which a transformation reduces the SIdV to the KdV equation. But there are special values of $\eps$ at which it comes close. First, the rational non-linearity of SIdV can be written as a polynomial while  retaining the advection-dispersion structure of the equation. To do so, we use the invariance of SIdV under rescaling to choose $u$ dimensionless and put $u = e^w$. Then\footnote{Of course, if we restrict to real $w$, this equation will apply to solutions where $u$ remains everywhere positive.}
	\beq
	w_t + \left[ (2- \eps) w_x^2 + (2-3 \eps) w_{xx} \right] w_x = \eps w_{xxx}.
	\label{e:variable-dispersion-w-form}
	\eeq
This polynomial form of SIdV was convenient for numerical evolution and also in our search for a Lagrangian. It also indicates that $\eps = 2, \frac{2}{3}$ are somewhat special. At these values, we get KdV-like dispersive wave equations with advecting velocities $\propto w_x^2$ and $w_{xx}$. These are among the simplest PT symmetric advecting velocities beyond KdV. Moreover, at $\eps = 2/3$, the sign of the `local diffusivity' is reversed. We see qualitative effects of this reversal in the stability of our numerical simulations as $\eps$ is decreased below $2/3$. What is more, at $\eps = 2/3$ SIdV is reducible to the modified KdV (mKdV) equation. Differentiating in $x$ at $\eps = 2/3$, putting $v = w_x$ and letting $x \to -x$ and $t \to \frac{3}{2} t$, we get the defocusing mKdV equation, which is integrable and related to KdV via the Miura transform \cite{drazin-johnson}
	\beq
	v_t - 6 v^2 v_x + v_{3x} = 0.
	\eeq

\section{Some similarity and travelling wave solutions of SIdV}
\label{s:solutions-of-SIdV}

\subsection{Similarity solutions}
\label{s:similarity}

SIdV is invariant under\footnote{More generally, SIdV is invariant under $x \to \lambda x, t \to \lambda^3 t, u \to \lambda^{\gamma} u$ for any $\gamma$. So $t^{-\gamma/3}u(x,t) = f(z)$ is a scale invariant combination. Since $\gamma$ is arbitrary, we restrict here to the simplest case $\gamma=0$.} $x \to \lambda x$ and $t \to \lambda^3 t$. So we seek solutions $u(x,t) = f(z)$ in the similarity variable $z = x^3/54t$. We get a third order non-linear ODE for $f(z)$
	\beq
	\frac{\eps z^2}{2} f''' f - z^2 f'' f' + \eps z f'' f
	- \frac{2 z}{3} f'^2 + \left(\frac{\eps}{9} + z \right) f'f = 0.
	\label{e:similarity-ode-3rd-order}
	\eeq
By the substitution $g = f'/f$ we reduce this to a cubic $2^{\rm nd}$ order ODE with variable coefficients
	\beq
	\frac{\eps z^2}{2} g'' + \left( \frac{3\eps}{2} -1 \right) z^2 gg'
	+ \eps z g' + \left( \frac{\eps}{2} -1 \right) z^2 g^3 + \left( \eps - \frac{2}{3} \right) z g^2 + \left( \frac{\eps}{9} + z \right) g = 0.
	\label{e:similarity-ode-2nd-order}
	\eeq
We haven't solved the similarity ODE in general, but in the dispersionless limit $\eps=0$, it becomes a {\em linear} ODE with a regular singularity at $z = 0$
	\beq
	z f'' + \frac{2}{3} f' - f = 0.
	\eeq
The two linearly independent solutions may be expressed in terms of the confluent hypergeometric function $_0F_1(a,z)$ or the modified Bessel function of the $1^{\rm st}$ kind $I_n(z)$:
	\beqs
	f_1(z) = \: _0F_1 \left(\frac{2}{3},z \right)
	&=& 1 + \frac{3z}{2} + \ldots \;
	\; = \;
	\Gamma\left(\frac{2}{3} \right) z^{\ov{6}} I_{-\ov{3}}\left(2\sqrt{z} \right)
	\quad {\rm and} \cr
	f_2(z) = \: z^{\ov{3}} \; _0F_1 \left(\frac{4}{3},z \right)
	&=& z^{\ov{3}} \left(1 + \frac{3z}{4} + \ldots \right) \;
	\; = \;
	\Gamma\left(\frac{4}{3} \right) z^{\ov{6}} I_{\ov{3}}\left(2\sqrt{z} \right).
	\eeqs
Both $f_1$ and $f_2$ are monotonic and grow $\propto e^{2\sqrt{z}}$ as $z \to \infty$. So they are bounded at late times ($z \to 0$ or $t \gg x^3$) but are unbounded at early times ($z \to \infty$ or $x^3 \gg t \to 0$). Interestingly, there is a unique (up to scale) linear combination $f(z) = \Gamma(4/3) \: f_1(z) - \Gamma(2/3) \: f_2(z)$ that is bounded for all $z \geq 0$ ($x,t \geq 0$). It begins at $f(0) = \Gamma(4/3)$ and monotonically decays to zero as $z \to \infty$.

\subsection{Travelling waves}
\label{s:trav-waves}

Here we discuss travelling waves ($u=f(\xi)$ with $\xi = x-ct$) for SIdV (\ref{e:sidv-eps}) on the unbounded domain $-\infty < \xi < \infty$. Travelling waves must satisfy the third order non-linear ODE $-cff' + 2 f' f'' - \eps f f''' = 0$. We may write this in `conservation law' form
	\beq
	-\frac{c}{2} \left( f^2 \right)' + \frac{2+\eps}{2} \left(f'^2 \right)' - \eps (ff'')' = 0
	\eeq
and integrate once to get
	\beq
	2 \eps f f'' - (\eps + 2) f'^2 + cf^2 + 3B = 0.
	\label{e:trav-wave-eqn-epsilon}
	\eeq
The substitutions $p = f'$ and $F = p^2$ give us a first order {\em linear} ODE for $F(f)$:
	\beq
	\eps f F'(f) - (\eps + 2) F + c f^2 + 3B = 0
	\qquad \text{or} \qquad
	\eps F'(w) - (\eps + 2) F + c e^{2w} + 3B = 0,
	\eeq
where $f = e^{w}$. This inhomogeneous $1^{\rm st}$ order ODE is reduced to quadrature using the integrating factor $e^{-(1+2/\eps)w}$. For $\eps \ne 0$, in terms of $r(w) = e^{-(1+2/\eps)w} F(w)$, we get
	\beq
	\eps r' + 3B e^{-(1+2/\eps) w} + c e^{(1-2/\eps)w} = 0.
	\label{e:r-of-w-1st-order-ode}
	\eeq
For $\eps \ne 0, \pm 2$, we integrate to get
	\beq
	F(w) = A e^{\left(1+ \frac{2}{\eps} \right)w} + \frac{c \eps \: e^{2w}}{2-\eps} + \frac{3B}{\eps + 2}.
	\eeq
In other words, the equation for travelling waves has been reduced to quadrature
	\beq
	\int \text{d}\xi = \pm \int \frac{\text{d}f}{\sqrt{F(f)}}
	\quad \text{where} \quad
	F = \begin{cases} 
	\frac{c}{2} f^2 + \frac{3 B}{2} & \text{if $\eps = 0$.} \\
	\half Af^2 - cf^2 \log f + \frac{3B}{4} & \text{if $\eps = 2$,} \\
	\frac{c}{4} f^2 + \frac{3B}{2} \log f - \half A & \text{if $\eps = -2$,  and} \\
	A f^{\left(1+\frac{2}{\eps} \right)} + \left(\frac{c\eps}{2-\eps}\right) f^2 + \frac{3B}{\eps+2} & \text{otherwise.}
	\end{cases}
	\label{e:trav-wave-eqn-epsilon-quadrature}
	\eeq
This travelling wave integral can be understood by a mechanical analogy \cite{drazin-johnson}. It is the zero `energy' condition $E = f'^2 + V(f) =0$ for the `coordinate' $f(\xi)$ at `time' $\xi$ of a non-relativistic particle of mass $2$ moving in the $1$-dimensional potential $V(f) = -F(f)$. Bounded travelling waves correspond to bound trajectories of this particle. Plotting $V(f)$ shows that, for appropriate ranges of $A,B$ and $c$, there are bounded spatially periodic/solitary travelling waves with heights between successive real simple/double zeros of $V(f)$. For generic $\eps$, the travelling wave integral $\int F^{-1/2} \: {\text{d}f}$ cannot be evaluated using elementary functions. But for $\eps = \infty, 0, -2/3$, the integral is trigonometric/exponential and for $\eps = 2/3, 1$ it is elliptic. 

To illustrate, we consider the case $\eps = 1$ where SIdV shares solitary wave solutions with KdV. Here we must evaluate an elliptic integral $\int (Af^3 + cf^2 + B)^{-1/2} \text{d}f$, and bounded travelling waves are (limits of) cnoidal waves. Suppose $A> 0$ and $F(f) = Af^3 + cf^2 + B = A(f-f_1) (f-f_2) (f-f_3)$ has three simple real zeros $0 \geq f_1 < f_2 < f_3 \geq 0$. Then we have a periodic cnoidal wave with trough at $f_1$ and crest at $f_2$, determined by
	\beq
	\pm (\xi - \xi_1) = \int_{f_1}^f \frac{\text{d}g}{\sqrt{A(g-f_1)(g-f_2)(g-f_3)}} \quad
	\text{where \;\; $f(\xi_1) = f_1$}.
	\label{e:integ-over-sqrt-cubic}
	\eeq
Transforming to $g = f_1 + (f_2 - f_1) \sin^2 \theta$ and defining the shape parameter $0 \leq m = \frac{f_2-f_1}{f_3-f_1} \leq 1$, (\ref{e:integ-over-sqrt-cubic}) becomes a standard incomplete elliptic integral of the first kind. It is inverted in terms of a Jacobi elliptic function ${\rm cn}(u;m) = \cos \phi$, with modulus $k$ ($m = k^2$)
	\beq
	\xi = \xi_1 \pm \frac{2 u }{\sqrt{A(f_3-f_1)}}
	\quad \text{where} \quad u = \int_0^\phi \frac{\text{d}\theta}{\sqrt{1-m \sin^2 \theta}}.
	\eeq
At the upper limit $g = f = f_2 - (f_2-f_1) \cos^2\phi$, so the cnoidal wave for $A > 0$ is
	\beq
	f = f_2 - (f_2 - f_1) \: \text{cn}^2 \left(\half \sqrt{A(f_3-f_1)} (\xi - \xi_1) \: ; \: m\right).
	\label{e:cnoidal-wave-A>0}
	\eeq
Its shape depends on $m$ while its wavelength and speed are
	\beq
	\lambda = \frac{4 K(m)}{\sqrt{A(f_3-f_1)}} \quad \text{and} \quad
	c = -A(f_1 + f_2 + f_3).
	\eeq
Here $K(m) = \int_0^{\pi/2} \frac{\text{d}\theta}{\sqrt{1-m \sin^2 \theta}}$ is the complete elliptic integral of the $1^{\rm st}$ kind.

The advecting velocity field $V = \frac{2 f''}{f}$ for cnoidal waves is finite since $f$ and $f''$ have common zeros. Moreover, by modifying the parameters $A,f_i$, we get nearby waves with the same feature. Unlike for KdV, where the shape, speed and wavelength of cnoidal waves are non-trivially modified upon a rescaling of amplitude, here $f \mapsto \lambda f$ produces a new cnoidal wave with the same $m$, $c$, $\lambda$ and phase $\xi_1$, since the constants transform as $(A,B, f_i) \mapsto (A/\lambda,\lambda^2 B, \lambda f_i)$.

If $A < 0$, the cnoidal wave extends between $f_2$ and $f_3$ and is given by
	\beq
	f(\xi) = f_2 + (f_3 - f_2) \; {\rm cn}^2 \left( \half \sqrt{A(f_1 - f_3)} \: (\xi - \xi_3) \, ; \, \tilde m \right) 
	\label{e:cnoidal-wave-A<0}
	\eeq
where $\tilde m = ({f_3 - f_2})/({f_3 - f_1})$ and $\lambda = {4K(\tilde m)}{\sqrt{A(f_1 - f_3)}}$.

Solitary waves are cnoidal waves of infinite wavelength. They occur when a pair of simple zeros of $F$ coalesce to form a double zero. For example, if $f_3 \to f_2$ in (\ref{e:cnoidal-wave-A>0}) holding $f_{1,2}$ and $A> 0$ fixed, then $m \to 1^-$, $K(m) \to \infty$ and we get a left-moving solitary wave of depression 
	\beq
	f(\xi) \to f_2 - (f_2 - f_1) \; \sech^2 \left( \half \sqrt{A(f_2 - f_1)} (\xi - \xi_1) \right) \quad \text{for} \quad A > 0.
	\label{e:sechsquare-wave-depr-non-zero-asymptote}
	\eeq
If the zeros $f_1 \to f_2$ coalesce in (\ref{e:cnoidal-wave-A<0}) we get a left-moving solitary wave of elevation
	\beq
	f(\xi) \to f_2 + (f_3 - f_2) \: \sech^2\left( \half \sqrt{A(f_2-f_3)} (\xi - \xi_3) \right) \quad \text{for} \quad A < 0.
	\eeq
Finally, if $A \to 0$, cnoidal waves reduce to sinusoidal waves with cubic dispersion,
	\beq
	f(\xi) = N \sin\left(\sqrt{-c} \left(\xi - \xi_0 \right) \right), \quad c < 0, \;\; N \;\; \text{arbitrary}.
	\eeq

\section{Search for a variational principle for SIdV}
\label{s:search-for-var-ppl}

To relate symmetries to integrals of motion, it is interesting to find a Lagrangian or Hamiltonian formulation for SIdV. The existence of a Lagrangian for a given equation depends on the field variables used and the sort of Lagrangian allowed. For instance, the dispersive term $u_{xxx}$ in KdV $u_t + 6 u u_x + u_{xxx} =0$  isn't the variation of any polynomial in $u$ and its derivatives. This is because every quadratic differential polynomial in $u$ involving three $x$-derivatives is a total derivative. Yet, as is well-known, if we put $u = \chi_x$, KdV follows from the Lagrangian density $\half \chi_t \chi_x + \chi_x^3 - \half \chi_{xx}^2$.

In looking for a Lagrangian for SIdV, we choose to work with $w = \log u$, which satisfies the KdV-like equation (\ref{e:variable-dispersion-w-form}) with polynomial non-linearity. By analogy with KdV, we put $w = \phi_x$ and seek a {\it polynomial} action in $\phi$ and its derivatives, whose Euler-Lagrange (EL) equations are
	\beq
	\phi_{xt} + \left[ (2- \eps) \phi_{xx}^2 + (2-3 \eps) \phi_{xxx} \right] \phi_{xx} = \eps \phi_{xxxx}.
	\label{e:SIdV-polynomial-phix}
	\eeq
$\phi$ is a natural variable since the linear part of the equation $\phi_{xt} = \eps \phi_{4x}$ admits the polynomial Lagrangian ${\cal L}_0 = \half \phi_t \phi_x + \frac{\eps}{2} \phi_{xx}^2$. We wish to add potentials $V,W$ to ${\cal L}_0$ to reproduce the quadratic $\phi_{xx} \phi_{3x}$ and cubic terms $\phi_{xx}^3$ in (\ref{e:SIdV-polynomial-phix}). To do so, we note a couple of general features. If $V$ is a monomial of degree $n$ in $\phi$ and its $x$-derivatives, then (1) the resulting terms in the EL equation form a differential polynomial of degree $n-1$, and (2) the total number of $x$-derivatives in each term of the differential polynomial are the same as in $V$. Therefore, to produce $\phi_{xx}^3$ in the EL equation, $W$ must be a quartic differential polynomial with $6$ $x$-derivatives, and to give $\phi_{xx} \phi_{3x}$, $V$ must be a cubic differential polynomial with $5$ derivatives. In other words,
	\beq
	V = a_1 \, \phi \phi \phi_{5x} + a_2 \, \phi \phi_x \phi_{4x} + a_3 \, \phi \phi_{2x} \phi_{3x} + a_4 \, \phi_x \phi_x \phi_{3x} + a_5 \, \phi_x \phi_{2x} \phi_{2x} = \sum_{i} a_i V_i.
	\eeq
Can $a_i$ be chosen so that the variation of $\int V \, \text{d}x$ gives $(2-3\eps)\phi_{xx} \phi_{3x}$? Unfortunately not, as\footnote{In particular, this means all the $V_i$ differ from one another by total $x$-derivatives.}
	\beq
	\frac{\delta}{\delta \phi(x)} \int V(\phi(y)) \:  \text{d}y \: = -\left( 10 a_1 - 5 a_2 + a_3 + 4 a_4 -2 a_5 \right) \left(2 \phi_{xx} \phi_{xxx} + \phi_x \phi_{4x} \right).
	\eeq
For no choice of $a_i$ can we produce just a quadratic monomial $\propto \phi_{xx} \phi_{xxx}$ in the EL equation. Similarly, we showed that there is no quartic differential polynomial $W$ that gives $\phi_{xx}^3$ upon variation. We conclude that there is no polynomial Lagrangian in $\phi$ and its derivatives leading to SIdV.

However, the polynomiality assumption is quite strong. There may be a non-polynomial Lagrangian in $\phi$ or one in a variable non-locally related to $\phi$. An interesting example of such a possibility occurs when $\eps = -\frac{2}{3}$ and SIdV becomes the linear dispersive equation $\rho_t + (2 / 3) \rho_{xxx} =0$ upon substituting $\rho = u^2$. This equation follows from ${\cal L} = \half \psi_t \psi_x - \ov{3} \psi_{xx}^2$, where $\psi_x = \rho$. However, $\psi = \int^x e^{2 \phi_y} \: \text{d}y$ is non-locally related to $\phi$, so the Lagrangian is non-local in $\phi$.

Another way around this negative result is that there may be a Hamiltonian that is a differential polynomial in $\phi$, but with non-canonical Poisson brackets. Such a  possibility is realized if $\eps = 2/3$, when SIdV can be transformed into the mKdV equation $v_t - 6 v^2 v_x + v_{3x} = 0$ by the substitution $v = \frac{u_x}{u}$ and a rescaling (\S \ref{s:SIdV-to-mKdV}). mKdV admits a Hamiltonian formulation $v_t = \{ H,v \}$ with
	\beq
	H = \half \int \left( v_x^2 + \frac{v^4}{6} \right) dx
	\quad \text{and} \quad \{ v(x) , v(y) \} = -\pdr_x \del(x-y).
	\eeq
So for $\eps = 2/3$, SIdV admits a polynomial Hamiltonian in the variable $v = (\log u)_x = w_x = \phi_{xx}$. It would be interesting to find a Hamiltonian/Lagrangian formulation of SIdV for other values of $\eps$.

\section{Numerical evolution of SIdV solitary waves: recurrent behavior}
\label{s:numerical-evolution-recurrence}

We numerically solved\footnote{The evolution was done using NDSolve on Mathematica. Stability of the numerical evolution was slightly enhanced by working with $w = \log u$ which satisfies $w_t = w_{3x} + w_x w_{xx} - w_x^3$. Positivity of  $u(x,0)$ was preserved at all times.} the SIdV initial value problem for $\eps = 1$ on the interval $[-\pi,\pi]$ with periodic boundary conditions. Numerical evolution of one $\sech^2$ wave produced a right-moving travelling wave, as expected from the exact solution on $(-\infty, \infty)$. We also considered two solitary waves\footnote{The sum over $j \in \mathbb{Z}$ ensures that the initial condition satisfies periodic boundary conditions $u(-\pi)=u(\pi)$. In practice, the sum was restricted to $|j| \leq 3$ since the remaining terms are exponentially small for $-\pi \leq x \leq \pi$.}
	\beq
	u(x,0) = \sum_{j} A_1 \sech^2\left(\frac{\sqrt{c_1}}{2}\left(x-x_1 + 2 \pi j \right)\right) + A_2 \sech^2\left(\frac{\sqrt{c_2}}{2} \left(x-x_2 + 2 \pi j \right) \right).
	\eeq
When the two solitary waves were initially separated by some distance but had the same heights and speed, they were observed to travel without much interaction, just like individual solitary waves. 

\begin{figure}
\label{f:2-sol-plots-1}
\subfigure[$u(x)$ at $t=0$, before, during and after collision of right-moving waves on $2\pi$-periodic interval.]{\includegraphics[width=7cm]{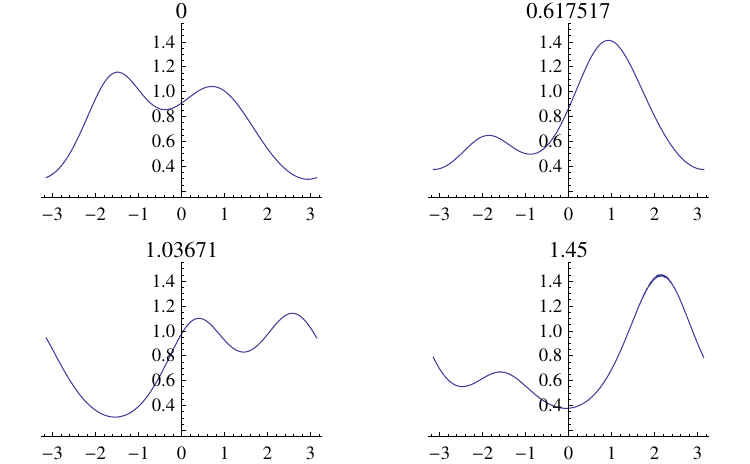} \label{f:uoft} }
\quad
\subfigure[$\int u^2 \; \text{d}x$ is conserved upto numerical errors, which cause it to grow by about $.3\%$ for $0 \leq t \leq 40$.]{\includegraphics[width=7cm]{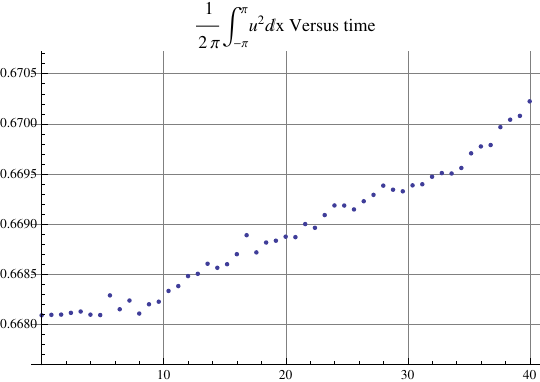}
\label{f:constancy-of-usquared}}
\caption{SIdV evolution of two solitary wave initial state. Times $0 \leq t \leq 40$ are indicated above the plots in fig. \ref{f:uoft}.}
\end{figure}

Next, we gave the two waves the same initial amplitudes $A_{1,2}= 1$ but different speeds (or widths) $(c_1, c_2) = (4,2)$ and locations $(x_1,x_2)=(-\pi/2, \pi/4)$ at $t = 0$ (fig. \ref{f:uoft}). So it would take each wave by itself a time of $\frac{\pi}{2}$ and $\pi$ to traverse the $2\pi$-interval. Solitary waves that decay at $\infty$ must be right-moving, so we couldn't give them opposing velocities. But $c_1 > c_2$, so wave-$1$ caught up with wave-$2$ due to periodic boundary conditions and collided with it from the rear. Then they separated into a small leading wave and a larger trailing wave. The original solitary waves did not retain their shapes. The smaller wave that emerged from the collision moved faster and caught up with the bigger one by going round the circle. During the next collision, the smaller wave rear-ended the larger one. The large one in front morphed into a fast small wave, leaving behind a slow large wave. This is illustrated in the last three plots of fig. \ref{f:uoft}. Qualitatively, this pattern seemed to repeat as the IVP was solved up to $t = 40$, allowing more than two dozen collisions to be observed\footnote{A maximum of $1200$ grid points were placed at an average spacing of $0.1\%$ the domain width $(2\pi)$. The qualitative features reported here were unchanged by adding $200$ more points. Over the time interval $0 \leq t \leq 40$, $1800$ nearly equal steps were taken with an average time step of $1.4\%$ of the time it took wave-$1$ in isolation to traverse the domain.}.

\begin{figure}
\label{f:2-sol-plots-2}
\subfigure[$\log_{10}|c_n(t)|^2$ for modes $0 \leq n \leq 5$ (top to bottom).]{\includegraphics[width=7cm]{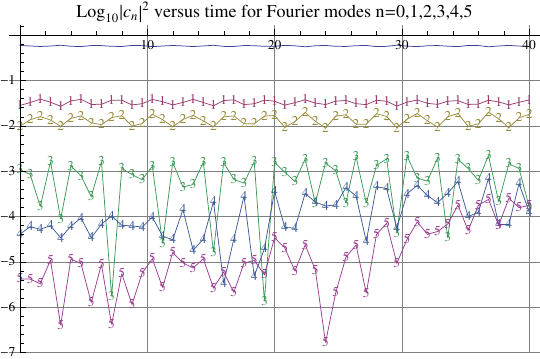} \label{f:fourier-coeffs-vs-time}} 
\quad
\subfigure[Rayleigh quotient from $5^{\rm th}$ order Fourier series]{\includegraphics[width=7cm]{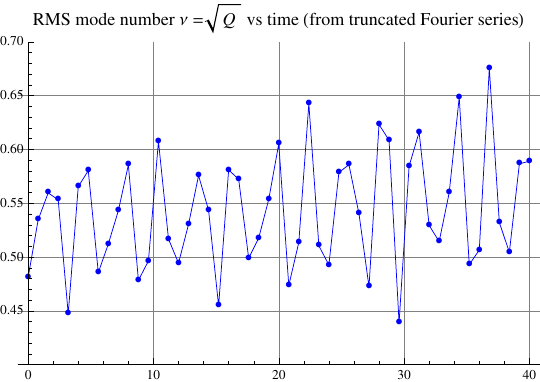} \label{f:rmsQ-2solitary-truncated}} 
\caption{ Time dependence of Fourier coefficients of $u(x)$ and gradient energy for a two-solitary-wave initial state.}
\end{figure}

Though not KdV solitons, the solitary waves that were involved in these collisions displayed a certain coherence, they did not dissipate nor degenerate into ripples. Despite not being periodic, the evolution seemed to approximately revisit earlier configurations. Interestingly, there was no equipartitioning of wave intensity. We illustrate this in fig. \ref{f:fourier-coeffs-vs-time} by plotting the absolute squares of the first few Fourier coefficients of $u(x)$ as a function of time\footnote{$c_n(t) = \ov{2\pi}\int u(x,t) \: e^{-inx} \: \text{d}x$, the negative coefficients $|c_{-n}|^2 = |c_n|^2$ contain no new information for real $u(x,t)$.}. There is some exchange of intensity among the first $3$ or so Fourier modes $c_{0,1,2}$, with an approximate periodicity of $T \approx 3$. But there is no appreciable leakage to higher Fourier modes $|c_{3,4,5}|^2$, which are uniformly three to six orders of magnitude smaller than $|c_0|^2$. There is some growth in $|c_{4,5}|^2$. But this can't be distinguished from an accumulation of numerical errors, which also caused the integral of motion $I = \int u^2 \; \text{d}x$ to increase by $0.3\%$ over a time $0 \leq t \leq 40$ (fig. \ref{f:constancy-of-usquared}). 

What is more, though the Raleigh quotient (`gradient energy' or mean square mode number)
	\beq
	Q(t) = \frac{\int |u_x(x,t)|^2 \; \text{d}x}{\int |u(x,t)|^2 \; \text{d}x} = \frac{\sum_n n^2 |c_n(t)|^2}{\sum_n |c_n(t)|^2}
	\eeq	
is {\em not} conserved, it seems to oscillate between bounded limits (fig \ref{f:rmsQ-2solitary-truncated}). The rms mode number hovers around $\nu = \sqrt{Q} \approx \half$. A calibration of $Q$ using the linear dispersive equation $u_t = u_{3x}$ indicates that there are about $2N+1 = \sqrt{12Q+1} \approx 2$ active degrees of freedom present, $c_0$ being the dominant one, with some contribution from $c_{\pm 1}$. These numerical simulations indicate that in a periodic domain, the SIdV equation displays recurrent behaviour (for $\epsilon=1$)  despite possessing only two (known) constants of motion.


\section{Advection-dispersion equations sharing KdV solitary waves}
\label{s:AD-eqns-with-KdV-solitary-waves}

One of the questions that puzzled us after the discovery of (\ref{e:AD-eqn}) by genetic programming, was whether there are other such non-linear advection-dispersion equations,
sharing the $\sech^2$ wave with KdV \cite{kudryashov2,kudryashov3,kudryashov4}. To explore this question let us consider the following generalized form of an advection dispersion
equation, 
	\beq
	u_t + V u_x = \del \: a \:  u_{xxx},
	\label{e:general-advection-dispersion-eqns}
	\eeq
 where $\delta$ is an arbitrary dimensionless parameter and $V(u,u_x\ldots)$ is an arbitrary function. We work in units where $a =1$ and assume the advecting velocity $V(u,u_x\ldots)$ to be translation-invariant, so that constants are generically solutions. We look for all $V$ and $\del$ for which (\ref{e:general-advection-dispersion-eqns}) admits every asymptotically decaying KdV travelling wave as a solution. We proceed by supposing that {\em every} decaying travelling wave $u(x,t)=f(x-ct) \equiv f(\xi)$ that solves (\ref{e:KdV-eqn}) also solves (\ref{e:general-advection-dispersion-eqns}) for the {\em same} speed $c$. Then
	\beq
	-c f' + 6 f f' + f''' = 0 \quad \text{and} \quad
	-c f' + V f' - \del f''' = 0.
	\eeq
Eliminating $f'''$ we get
	\beq
	\left\{V -(\del + 1)c + 6 \del f \right\} f' = 0.
	\label{e:condition-on-V-and-delta-with-c}
	\eeq
This equation for $V$ and $\del$ must be satisfied for arbitrary decaying KdV travelling waves. Since KdV admits non-constant travelling waves, $f' \nequiv 0$. So we must have $V = (\del + 1)c - 6 \del f$. $c$ is eliminated using the KdV equation $-c f' + 6 f f' + f''' =0$. However, simply substituting $c = 6f + f'''/f'$ gives $V_1 = 6u + (\del+1) \frac{u_{3x}}{u_x}$ leading back to the KdV equation. On the other hand, we could integrate  either once or twice while omitting integration constants for asymptotically decaying waves and find
	\beq
	-cf + 3 f^2 + f'' = 0 \quad \text{or} \quad
	-\frac{c}{2} f^2 + f^3 + \half (f')^2 = 0.
	\label{e:KdV-after-one-two-integ-decaying-case}
	\eeq
Eliminating $c$ from (\ref{e:condition-on-V-and-delta-with-c}), we find 
$\del$ is arbitrary and $V$ can take the functional forms
	\beq
	V_2 = 
	3(1-\del) u + (1 + \del) \frac{u_{xx}}{u} \quad  \text{or} \quad
	V_3 = 2(1-2\del) u + (1 + \del) \frac{u_x^2}{u^2}.
	\label{e:advecting-vel-sharing-KdV-sech-waves}
	\eeq
However, these are not the only possibilities. Instead of integrating $-c f' + 6 f f' + f''' =0$, we could differentiate this expression (any number of times!) and then use it to eliminate $c$. In this manner we get an infinite sequence of higher order advecting velocities for which (\ref{e:general-advection-dispersion-eqns}) admits (\ref{e:KdV-solitary-wave-rtmov-elevation}) as a solution:
	\beqs
	V_4 &=& 6u + (1 + \del) \left( \frac{6 u_{x}^2}{u_{xx}} 
	+ \frac{u_{4x}}{u_{2x}} \right), \qquad
	V_5 = 6u + (1 + \del) \left( \frac{18 u_{xx} u_x}{u_{3x}} 
	+ \frac{u_{5x}}{u_{3x}} \right),
	 \cr
	V_6 &=& 6u + (1 + \del) \left( \frac{24 u_{3x} u_x}{u_{4x}} 
	+ \frac{18 u_{xx}^2}{u_{4x}} + \frac{u_{6x}}{u_{4x}} \right), 
	\; \ldots
	\label{e:higher-order-advective-vel-sharing-sech-waves}
	\eeqs
$V_2$ and $V_3$ are distinguished in that they are linear combinations of $6u$ (KdV) and new scale-invariant advecting velocities $\frac{u_{xx}}{u}$ or $\frac{u_x^2}{u^2}$. $V_{n \geq 4}$ differ from KdV by non-scale-invariant advecting velocities. They also involve higher
order derivatives than in the dispersive term $u_{xxx}$.

Thus, KdV is at the center of an infinite dimensional space of advection-dispersion equations with $\sech^2$ solitary wave solutions. In the space of equations, we can go out a `distance' $1+\del$ in any of the directions defined by $V_n$ while retaining this solution. In the simplest (least non-linear) case $V_2$, we get the KdV-SIdV family interpolating between KdV ($\del = -1$) and (\ref{e:AD-eqn}) ($\del = 1$)
	\beq
	u_t + \left( 3(1-\del) u + (\del + 1) \frac{u_{xx}}{u} \right) u_x
	\; =  \; \del \: u_{xxx}.
	\label{e:KdV-SIdV-eqn}
	\eeq
	It may be noted that $\delta$ and $\epsilon$ are distinctly different parameters and we have therefore kept a separate notation. 
    $\epsilon$ is a dispersion coefficient that is exclusive to SIdV advection and  $\delta$ is a different parameter that allows the KdV and
    the SIdV(when $\epsilon =1$) equations to be regarded as part of the SIdV-KdV one parameter family.
Like KdV and SIdV, (\ref{e:KdV-SIdV-eqn}) preserves $\int u^2 \: \text{d}x$ since it can be written in conservation law form
	\beq
	\half \left(u^2 \right)_t + \left((1-\del) u^3 + \left(\half + \del \right) u_x^2 - \del uu_{xx} \right)_x = 0.
	\eeq
As we omitted a constant of integration in (\ref{e:KdV-after-one-two-integ-decaying-case}), the above argument {\em does not} guarantee that every other KdV travelling wave is a solution of (\ref{e:KdV-SIdV-eqn}). For example, there are non-decaying KdV travelling waves (e.g. $2 \sech^2(x+2t) -1$) which are not solutions of (\ref{e:KdV-SIdV-eqn}) when $\del = 1$. Nevertheless, (\ref{e:AD-eqn}) has cnoidal and non-decaying $\sech^2$ solitary waves of its own. Had we kept the constant of integration $-cf + 3f^2 + f'' = A$, we would have found $V = 3(1-\del) f + (\del + 1) \left(\frac{f''}{f} - \frac{A}{f} \right)$. By varying $A$ we get equations that share particular asymptotic classes of travelling wave solutions with KdV.

\section{Discussion and conclusion}
\label{s:discussion}

In this paper we studied a scale-invariant analogue of the KdV equation, $u_t + 2 u_{xx} u_x/u = \eps u_{xxx}$, which we named the SIdV equation. SIdV is one of the two simplest translation, scale and space-time parity-invariant non-linear advection-dispersion equations\footnote{The other one involves the reciprocal of SIdV's advecting velocity: $u_t + \frac{u}{u_{xx}} u_x = \pm u_{xxx}$.}. The dimensionless parameter $\eps$ measures the strength of dispersion relative to advection. For $\eps = 1$, SIdV shares the $\sech^2$ solitary wave with KdV, as originally discovered by genetic programming. When $\eps = \pm 2/3,0$, the equation reduces to the integrable mKdV equation, to a linear dispersive wave equation and to a non-linear diffusion equation. Like KdV, SIdV admits similarity solutions, solitary and cnoidal travelling waves, and remarkably, even plane waves. SIdV may be written in conservation law form in two ways (\ref{e:cons-law-form-sidv}, \ref{e:cons-law-form-sidv-second}), leading to two integrals of motions. In general, we have not been able to find a Lagrangian for SIdV, indeed we could show that there is no Lagrangian polynomial in $\phi$, where $\phi_x = \log u$. However, we found a Lagrangian in a different variable at $\eps = -2/3$ and a Hamiltonian when $\eps = 2/3$.

For $\eps = 1$, numerical evolution of a pair of solitary waves did not show KdV-like soliton scattering, nor did the wave intensity tend towards equipartition among all Fourier modes, even after three dozen collisions. The wave intensity appeared to circulate among the modes present in the initial state. Effectively, a finite number of degrees of freedom appeared to take part in the dynamics. This was also manifested in the boundedness of the gradient energy $Q$ in our simulations. So despite the apparent absence of soliton scattering and presence of only two known conserved quantities, (\ref{e:AD-eqn}) seems to display `Birkhoff recurrence' like the Fermi-Pasta-Ulam or KdV systems. Thus, it may provide a counter example to the idea that integrability is necessary for recurrence. This was suggested in \cite{thyagaraja1979}, where the number of effective degrees of freedom of a conservative nonlinear wave equation was identified as the possible origin of a Birkhoff recurrence, a concept generalizing the well-known Poincar\'{e} recurrence of finite-dimensional Hamiltonian systems. The above-mentioned SIdV-like equation with reciprocal advecting velocity $V = u/u_{xx}$ is also interesting from this viewpoint, as it admits the conserved quantity $\int u_x^2 \: \text{d}x$. It follows by Poincare's inequality that the mean square mode number $Q$ is bounded, and we expect it too to display recurrence!

There are of course several other interesting directions such as perturbation theory for (\ref{e:KdV-SIdV-eqn}) around the KdV limit, and for SIdV around the solvable $\eps = \pm 2/3$ limits. Additional conserved quantities and solutions on bounded domains with homogeneous boundary conditions or on a half-line are also of interest. More generally, we would like to study structural issues like regularity and stability. We are also examining multi-dimensional analogues of SIdV as well as a complex version $({\rm i} \psi_t + \frac{\psi_{xx}}{\psi^*} \psi_x = {\rm i} \psi_{xxx})$ which has a locally conserved `probability' density $|\psi|^2$. 

Motivated by the computer-aided discovery of (\ref{e:AD-eqn}) we searched for other advection-dispersion equations $u_t + V u_x = \delta u_{xxx}$ sharing the KdV solitary wave. We found that there is an infinite sequence of 1-parameter families of advective velocities $V_n(u;\del)$ (\ref{e:advecting-vel-sharing-KdV-sech-waves}, \ref{e:higher-order-advective-vel-sharing-sech-waves}) that generalize the KdV equation while retaining its $\sech^2$ solitary wave solution. They involve a linear combination of KdV-like $V \propto u$ and new rational advective velocities. The first two of these have the feature of being scale-invariant: $V \propto u_{xx}/u$ (SIdV) and $u_x^2/u^2$. Among all these equations sharing KdV solitary waves, the KdV-SIdV
family (43) appears to be unique in not involving higher order derivatives and admitting a positive definite conserved density ($u^2$). It would be interesting to study this generalization in
greater detail.

{\flushleft \bf Acknowledgements:} The work of GSK was supported by a Ramanujan grant of the Dept. of Science and Technology, Govt. of India. 

\appendix
\section{Computer-aided discovery of SIdV equation}
\label{a:GP-sniffer}

The basic idea behind Genetic Programming (GP) is to simulate a stochastic process by which genetic traits evolve in offspring, through a random combination of the genes of the parents. Following the seminal work by Koza \cite{koza}, the GP framework provides a very useful stochastic engine to discover various solution regimes in a complex search terrain of a given problem. It is known that an evolutionary method is especially useful when direct methods are not available. In order to set up a GP engine, a non-linear chromosome structure representing a candidate solution is set up that can potentially grow to a true solution by successive applications of GP operators of selection, copy, crossover and mutation. The quality of a given chromosome is defined and scaled down typically to a fitness range $[0, 1]$ with fitness $1$ signifying a true solution. Stochastically generated chromosomes fill an initial pool that is evolved through successive generations in which potentially strong candidate chromosomes are selected based on their fitness values. They undergo possible refinements through GP operators and hopefully march towards a solution with fitness=1. In using GP to deduce a PDE like the KdV (in symbolic form) from an analytic travelling wave solution, one begins by considering a general expression for a third order ODE,
	\beq
	f'''(\xi) = C\left({\xi, ~f(\xi), ~f'(\xi), ~f''(\xi) } \right)
	\label{Eqn:ODEEquation}
	\eeq
where $f(\xi)$ is a function of the travelling wave phase $\xi = x - ct$. For example, a chromosome during a GP iteration could be $C = 1.1 f'' + 2 f' (f' - 3 f) + \xi$. The fitness parameter is then estimated at each stage by examining the mean-squared difference between the `chromosomal value' of $f'''$ and its value at the given analytic solution. GP follows a `fitness driven evolution path' by minimizing the error in admitting the given function as a solution. We carried out a number of GP experiments for the ${\rm sech}^2$ KdV solitary wave (\ref{e:KdV-solitary-wave-rtmov-elevation}). Due to a stochastic search procedure adopted by GP, it was found to be too slow. We improved and accelerated it by introducing a sniffer technique \cite{abhijit-dilip} that carried out a local search at regular intervals to enhance the minimization procedure. Our improved  GP approach was quite successful in inferring PDEs. Starting with the KdV solitary wave and its derivatives, the method not only reproduces the KdV equation, but also gives the $\text{SIdV}$ equation (\ref{e:AD-eqn}).

\small


\end{document}